\documentclass{iopart}
\usepackage{graphicx}
\usepackage{psfrag}
\usepackage{amssymb}
\usepackage[square,comma,sort&compress]{natbib}
\usepackage{bm}
\def\l{\langle}

\def\r{\rangle}

\begin{document}

\title[Spin and chiral stiffness of the 2D {\em XY\/} spin glass]
{Spin and chiral stiffness of the {\em XY\/} spin glass in two dimensions}
\author{M Weigel\dag\ and M J P Gingras\ddag}
\address{\dag Department of Mathematics, School of Mathematical and Computer
  Sciences, Heriot-Watt University, Edinburgh, EH14~4AS, UK }
\address{\ddag Department of Physics, University of Waterloo, Waterloo, Ontario,
  N2L~3G1, Canada}
\eads{\mailto{M.Weigel@hw.ac.uk}, \mailto{gingras@gandalf.uwaterloo.ca}}

\begin{abstract}
  We analyze the zero-temperature behavior of the {\em XY\/} Edwards-Anderson spin
  glass model on a square lattice. A newly developed algorithm combining exact
  ground-state computations for Ising variables embedded into the planar spins with a
  specially tailored evolutionary method, resulting in the {\em genetic embedded
    matching\/} (GEM) approach, allows for the computation of numerically exact
  ground states for relatively large systems. This enables a thorough
  re-investigation of the long-standing questions of (i) extensive degeneracy of the
  ground state and (ii) a possible decoupling of spin and chiral degrees of freedom
  in such systems. The new algorithm together with appropriate choices for the
  considered sets of boundary conditions and finite-size scaling techniques
  allows for a consistent determination of the spin and chiral stiffness scaling
  exponents.
\end{abstract}

%75.50.Lk Spin glasses and other random magnets
%64.60.Fr Equilibrium properties near critical points, critical exponents
%02.60.Pn Numerical optimization
\submitto{\JPCM}

\section{Introduction}

With their rich behavior at low temperatures, spin glasses take a prominent role in
the large class of magnetic systems with frustration.  The most commonly considered
Hamiltonian is that of the Edwards-Anderson (EA) model \cite{kawashima:03a},
\begin{equation}
  \label{eq:EA_model}
  {\cal H} = -\sum_{\l ij\r}J_{ij}\,\bm{S}_i\cdot\bm{S}_j
\end{equation}
with O($n$) spins $\bm{S}_i$ and random, nearest-neighbor couplings $J_{ij}$. The
wealth of behavior of these systems is attributed to the random disorder augmenting
the frustration effects. Unfortunately, it is precisely this {\em quenched\/}
disorder that provides an exceptional challenge for the application of the various
analytical approximation methods well known from the treatment of homogeneous
systems. Owing to these difficulties, most of the advances in the understanding of
spin glass systems beyond the celebrated mean-field theory of the
Sherrington-Kirkpatrick model \cite{sk} have been on account of ever more
sophisticated numerical simulation and optimization techniques \cite{kawashima:03a}.
For two-dimensional (2D) systems, where for short-range interactions spin glass order
is restricted to zero temperature, the investigation of ground states is of prominent
interest. In general, finding (exact) ground states of spin glass models is a
computationally hard problem, where the amount of computer time grows exponentially
with the size of the system \cite{hartmann:book}.  Here, we explore a new avenue to
advance methods for the so far much less investigated case of systems with {\em
  continuous spins\/}: we introduce a novel approximate optimization algorithm which,
for the 2D {\em XY\/} spin glass discussed here, allows to find numerically exact
ground states with good confidence for systems of up to about $30 \times 30$ spins
\cite{weigel:05f,weigel:prep}.
 
Generalizing Peierls' argument for the stability of the ordered phase in homogeneous
systems to situations with quenched disorder, a droplet scaling theory for spin
glasses has been formulated \cite{bray:87a}. Therein, the role of the droplet surface
(free) energy is taken on by the {\em width\/} $J(L)$ of the distribution of random
couplings for a real-space renormalization group decimation at length scale $L$. In
the course of renormalization, $J(L)$ scales as $J(L) \sim L^{\theta_s}$, defining
the {\em spin stiffness exponent\/} $\theta_s$. If the system scales to weak
coupling, $\theta_s < 0$, spin-glass order is unstable at finite temperature and the
system is below its lower critical dimension. This is the situation for the EA model
in 2D \cite{kawashima:03a}, where then $\theta_s$ describes the properties of the
{\em critical\/} point at temperature $T=0$, relating the correlation length exponent
$\nu = -1/\theta_s$ \cite{bray:87a}. Numerically, the domain-wall free energy might
be determined from the energy difference between ground states of systems with
different types of boundary conditions (BCs) chosen such as to induce a relative
domain wall \cite{bray:87a}. For the $n=1$ Ising spin glass, the ground-state problem
on planar graphs is an exception to the rule, being polynomial computationally
\cite{hartmann:book}.  Hence, large systems can be treated, leading to reliable
estimates of $\theta_s = -0.282(2)$ (Gaussian $J_{ij}$) resp.\ $\theta_s = 0$
(bimodal $J_{ij}$) \cite{hartmann:01a,kawashima:03a}. Due to the presence of strong
finite-size corrections, relatively large system sizes and/or elaborate finite-size
scaling techniques appeared mandatory for consistent estimates of
$\theta_s$ \cite{hartmann:01a,carter:02a}. However, for the case $n>1$ of continuous
spins, which is more relevant to real materials, the lack of effective and efficient
algorithms for finding exact ground states and the necessary restriction to small
systems with $L\le 12$ have led to rather less consistent estimates, moving in the
range $\theta_s \in [-1,-0.75]$ \cite{morris:86a,kawamura:91,maucourt:98a}.

Moreover, the increased symmetry of the order parameter in the continuous spin case
has led to speculations about a decoupling of spin and chiral variables
\cite{kawamura:87a}: since the pattern of frozen spins in the glassy phase has
internal degrees of freedom, there is a factual difference between proper and
improper rotations expressed in the decomposition $\mathrm{O}(n) = \mathrm{SO}(n)
\times \mathbb{Z}_2$ \cite{villain:77a}. The additional Ising like chirality
variables might order independently of the spins (for systems above their lower
critical dimension) or, at least, show a different stiffness against fluctuations,
resulting in a scaling exponent $\theta_c$ possibly distinct from $\theta_s$. Indeed,
measurements of the chiral stiffness for small systems yielded $\theta_c \approx
-0.38$ \cite{kawamura:91,maucourt:98a}, different from $\theta_s$ above. More
recently, however, Kosterlitz and Akino \cite{kosterlitz:99a} argued that the choice
of BCs in previous studies was flawed and they suggest a possibly more appropriate
approach leading to $\theta_s \approx -0.38 \approx \theta_c$ again for sizes $L \le
10$. The hardly compatible previous results for this system hence raise several
methodological questions: have numerically exact ground states been found?, are the
apparent strong finite-size effects under control?, have the considered sets of BCs
been chosen such as to properly select the intended excitations?, and what is the
role of scaling corrections explicitly depending on these BCs?

\section{Genetic embedded matching approach}

The treatment of large samples for the 2D Ising case is enabled by a transformation
to an equivalent problem on the complete graph of frustrated plaquettes: following
their definition, for each spin configuration frustrated plaquettes have an odd
number of broken bonds around their perimeter, whereas unfrustrated plaquettes have
an even number of broken bonds. Thus, drawing ``energy strings'' dual to the broken
bonds, these connect pairs of frustrated plaquettes, and the total energy of
(\ref{eq:EA_model}) is (up to a constant) given by the total length of energy
strings, such that the ground state corresponds to a {\em minimum-weight perfect
  matching\/} of frustrated plaquettes \cite{hartmann:book}. The
matching problem can be solved in polynomial time using Edmonds' algorithm
\cite{edmonds:65a}, and for the case of planar graphs its solution is guaranteed to
transform back to a valid spin configuration \cite{bieche:80a}. This does not
directly apply to the continuous spins considered here. We suggest, however, to {\em
  embed\/} Ising variables into the planar spins by decomposing $\bm{S}_i =
\bm{S}_i^\parallel + \bm{S}_i^\perp = (\bm{S}_i\cdot\bm{r})\bm{r} + \bm{S}_i^\perp$
relative to a random direction $\bm{r}$ in spin space. With respect to reflections of
spins along the direction $\bm{r}$, the Hamiltonian (\ref{eq:EA_model}) decomposes as
${\cal H} = {\cal H}^{r,\parallel} + {\cal H}^{r,\perp}$ with ${\cal H}^{r,\parallel}
= -\sum_{\langle i,j\rangle}\tilde{J^r_{ij}}\,\epsilon_i^r \epsilon_j^r$, and
\begin{equation}
  \label{eq:embedd}
  \tilde{J_{ij}^r} = J_{ij}|\bm{S}_i\cdot\bm{r}||\bm{S}_j\cdot\bm{r}|,\;\;\;
  \epsilon_i^r = \mathrm{sign}(\bm{S}_i\cdot\bm{r}).
\end{equation}
Hence, for any fixed $\bm{r}$ and restricting the movement of spins to reflections
along $\bm{r}$, the Hamiltonian (\ref{eq:EA_model}) for arbitrary $n>1$ takes on the
form of an Ising model. Consequently, Edmonds' algorithm can be applied to find (one
of) the ground state(s) of the embedded Ising model. It is obvious that this can
never increase the energy of the full Hamiltonian (\ref{eq:EA_model}), but the state
found depends on the choice of random direction $\bm{r}$. To statistically
recover the full O($n$) symmetry of the Hamiltonian, a series of subsequent
minimizations is performed with respect to successive random choices of $\bm{r}$,
thus gradually decreasing the total energy via non-local moves. We refer to this
approach as {\em embedded matching\/} technique.

It can be shown that, although when the full Hamiltonian (\ref{eq:EA_model}) is in a
ground state, all embedded Ising Hamiltonians ${\cal H}^{r,\parallel}$ must be in one
of their respective ground states as well, successive minimizations with respect to
random directions $\bm{r}$ are not guaranteed to drive the system towards its
absolute energy minimum. In other words, the non-local embedded matching dynamics has
metastable states, but many less than the simple local spin quench dynamics used
before \cite{morris:86a,maucourt:98a}. To converge to ground states with high
probability, we insert the embedded matching technique as minimization component
(``subroutine'') in a genetic algorithm \cite{pal:96a}: we consider a whole
population of candidate ground-state configurations and simulate an evolutionary
development by re-combining (or {\em crossing over\/}) neighboring pairs of parent
configurations followed by minimization runs for the resulting offspring and
replacement of the parents in case of lower energy of the offspring. In analogy with
the approach of Ref.\ \cite{pal:96a}, the crossover is performed in a ``triadic''
fashion, comparing the overlaps with a third, reference configuration. This layout is
complemented by intermittent random mutation steps and performance-guided halving of
the population at certain stages to find an optimum balance between ``genetic''
diversity and efficiency of optimization \cite{pal:96a}. The choice of operation for
the crossover of configurations is found to be crucial for the efficiency of the
approach: it turns out that a random exchange of suitably defined rigid spin clusters
is far more efficient than an exchange of single spins. These clusters denote regions
which are highly optimized inside for all configurations of the population (i.e.,
metastable states), but which have to undergo a series of independent rigid O($n$)
transformations to make up a true ground state of the system. Careful choice of the
parameters of the resulting {\em genetic embedded matching\/} (GEM) algorithm and
application of various statistical tests ensure that indeed independent runs for a
single given realization of the disorder variables $J_{ij}$ always converge to a
state of the same energy, up to unprecedented machine precision, which in this way
can be guaranteed to be a ground state with high reliability
\cite{weigel:05f,weigel:prep}.

We here concentrate on the symmetric, bimodal $\pm J$ distribution.  For this case we
find that the ground states computed in independent runs for a single disorder
realization are always identical to each other up to a global O($n$) transformation,
indicating the lack of accidental degeneracies in contrast to what is found for the
bimodal Ising case.  Hence, after averaging over disorder, the ground state is
ordered and the ground-state spin correlation function is constant, implying $\eta =
0$. To determine the asymptotic ground-state energy per spin $e_\infty$,
ground-states were computed for $L\times L$ square-lattice systems with $L=6$, 8, 10,
12, 16, 20, 24, and 28 for open and open-periodic BCs and $5\,000$ disorder replica
per size.  Finite-size corrections are expected to be purely analytic for the case of
open BCs \cite{hartmann:04a}, and a fit to the ansatz $e(L) =
e_\infty+a/L+b/L^2+c/L^3$ yields $e_\infty = -1.5520(14)$ with quality-of-fit
$Q=0.35$. For the open-periodic case, an additional non-analytic term $\propto
L^{-(d-\theta)}$ is expected to occur \cite{hartmann:04a}, and a fit of the
corresponding data to the form $e(L) = e_\infty+a/L+b/L^{2-\theta}$ gives $e_\infty =
-1.5525(13)$, $\theta=-0.49(69)$, $Q=0.35$, perfectly consistent with the
open-boundary result for $e_\infty$ and, due to the large statistical error, only in
qualitative agreement with the expected value for the spin-stiffness exponent
$\theta$. The resulting $e_\infty$ is about $10\%$ lower than the value
$e_\infty=-1.402$ of the bimodal Ising spin-glass \cite{hartmann:04a}.

\section{Spin and chiral stiffness exponents}

\begin{figure}[tb]
  \centering
  \includegraphics[clip=true,keepaspectratio=true,width=5.5cm]{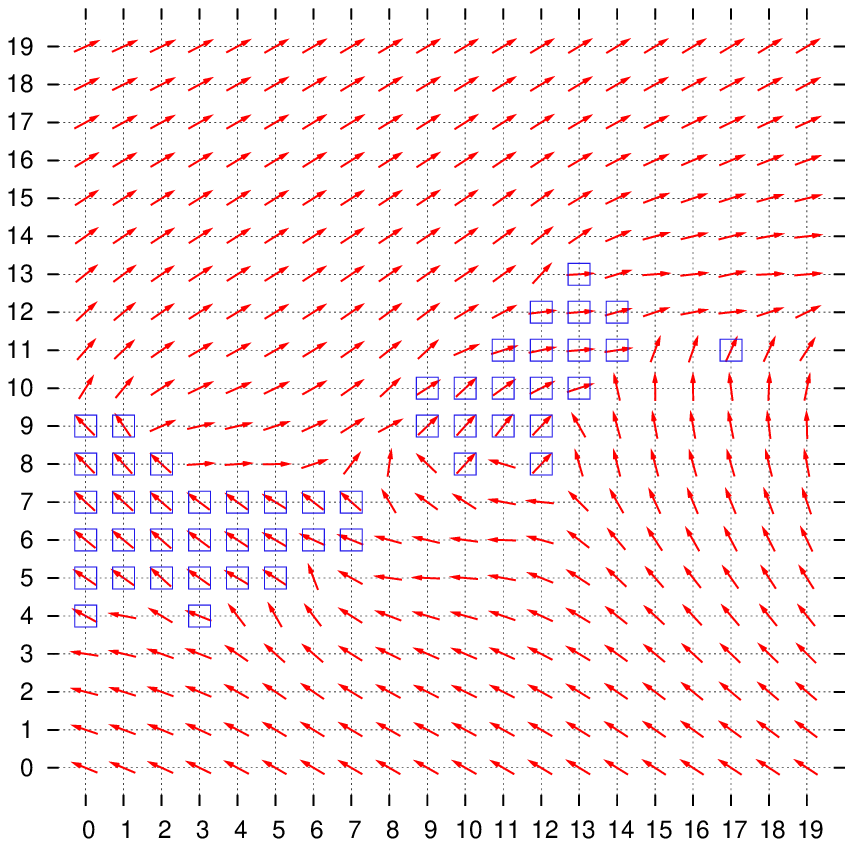}
  \hspace*{0.2cm}
  \includegraphics[clip=true,keepaspectratio=true,width=5.5cm]{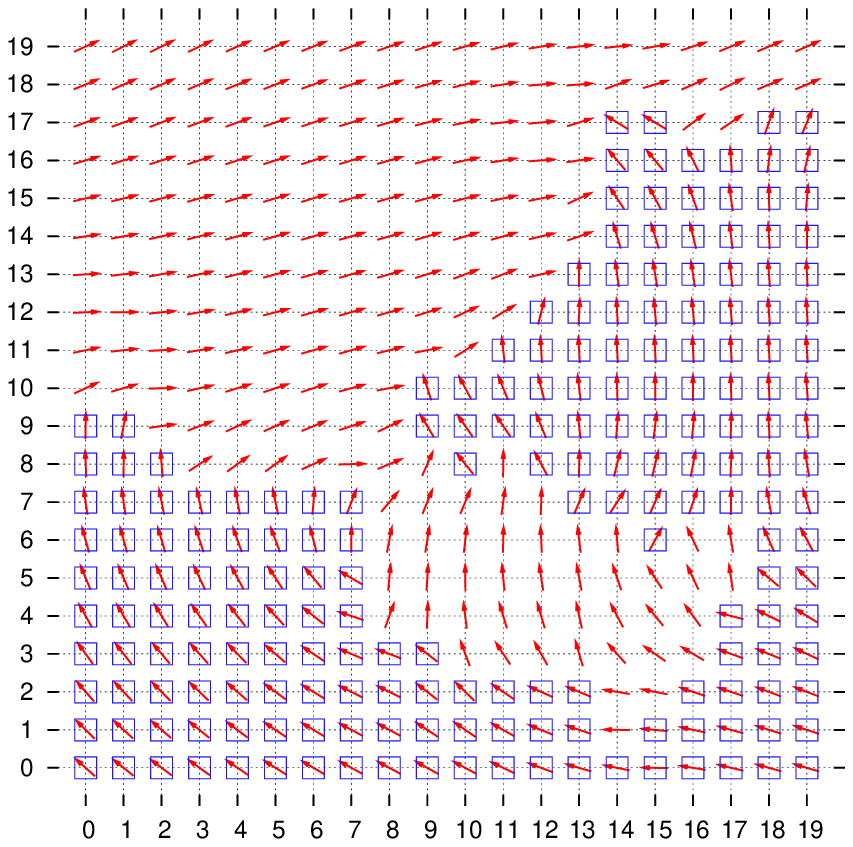}
  \caption
  {Local rotation matrices between the ground states for a single $20 \times 20$
    disorder realization with open boundaries relative to domain-wall BCs for spin
    (left) and chiral (right) excitations.}
  \label{fig:snapshots}
\end{figure}

Conventionally, domain-wall energies have been measured by comparing ground states
for periodic and antiperiodic (P/AP) BCs \cite{morris:86a,kawamura:91,maucourt:98a}.
In Ref.~\cite{kosterlitz:99a} it was argued, however, that the periodicity in both
types of BCs forces domain walls into the system, such that the corresponding energy
difference might not properly capture the energy of a single domain wall. There, an
improvement is suggested by optimizing over an additional global twist variable along
the boundary under consideration. Here, to start with the cleanest possible setup, in
addition to the conventional P/AP BC set, we consider open and domain-wall (O/DW)
BCs, where for the latter the relative orientations of spins linked across the
boundary are either tilted by an angle $\pi$ for spin domain walls or reflected along
an arbitrary but fixed axis for chiral domain walls by the introduction of very
strong bonds \cite{hartmann:01a,weigel:prep}. In Figure \ref{fig:snapshots} we show
snapshots of spin and chiral excitations forced into the system by the O/DW BCs. To
this end, we computed a locally averaged O($2$) rotation matrix relating the
configurations with O and DW BCs and translated it back into a rotation angle (the
arrows) and the sign of the determinant ($-1$ for the blue squares). It is apparent
that in contrast to the discrete Ising case the spin domain walls are rather smeared
out and that to a certain extent the system appears to relax the spin excitation also
through the chiral mode if it is found to be softer locally (and vice versa for the
chiral excitation).

\begin{figure}[tb]
  \centering
  \includegraphics[clip=true,keepaspectratio=true,width=7cm]{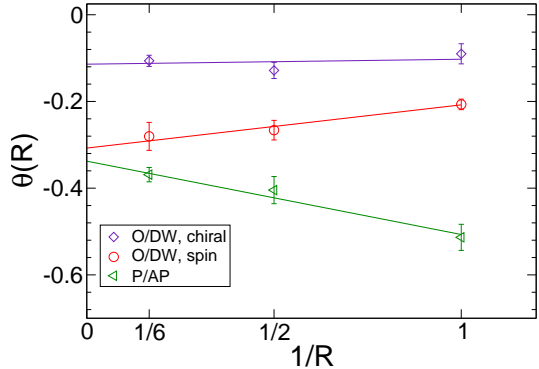}
  \caption
  {Aspect-ratio scaling of the stiffness exponents $\theta_s$ and $\theta_c$ for
    aspect ratios $R=1$, $2$ and $6$ as a function of $1/R$.}
  \label{fig:ASR_scaling}
\end{figure}

From the scaling of the domain-wall energies, $[|\Delta E|]_J\sim L^\theta$, we find
a strong crossover for the P/AP data from $\theta_s= - 0.724(21)$ for $L \le 12$ to
$\theta_s = -0.433(26)$ for $L\ge 16$, indicating large finite-size effects and a
movement from the value found for small P/AP computations in previous works
\cite{morris:86a,kawamura:91,maucourt:98a} towards the ``optimum twist'' value of
Ref.~\cite{kosterlitz:99a}. The O/DW BCs, on the other hand, yield $\theta_s =
-0.207(12)$ for spin excitations resp.\ $\theta_c = -0.090(23)$ for the chiral domain
walls.  Hence, although it is already clear that the true stiffness exponents are
much less negative than estimated before, there is still a sizable difference between
the P/AP and O/DW results for $\theta_s$, indicating incomplete control over
finite-size effects. To improve on this, we take into account that, due to the
independence of BCs for systems in one dimension, corrections depending on BCs should
disappear as more and more elongated systems are being considered \cite{carter:02a}.
Thus, we additionally performed computations for $L\times M$ systems (the change of
BCs happening along the edges of length $L$) with aspect ratios $R\equiv M/L = 2$ and
$6$ with the same statistics. The results are presented in Figure
\ref{fig:ASR_scaling} for the case of P/AP and O/DW BCs, respectively. Guided by the
experience from the Ising case, we expect corrections depending on BCs to disappear
as $\theta(R) = \theta(R\!=\!\infty)+A_R/R$ for large $R$, and indeed the P/AP and
O/DW spin data seem to converge for large $R$, a fit to the given form yielding
$\theta_s(R\!=\!\infty)=-0.338(20)$ for P/AP BCs and
$\theta_s(R\!=\!\infty)=-0.308(30)$ for O/DW BCs. The O/DW chiral data, on the other
hand, give $\theta_c(R=\infty)=-0.114(16)$, clearly distinct from $\theta_s$.

\section{Conclusions}

Employing a novel ``genetic embedded matching algorithm'', we computed numerically
exact ground states for the 2D {\em XY\/} EA spin glass with $\pm J$ couplings and up
to $28 \times 28$ spins. No accidental degeneracies occur, implying $\eta = 0$.
Analyzes of the domain-wall energies are hampered by strong finite-size effects
which, however, can be controlled using the aspect-ratio scaling technique. We find
consistent estimates of $\theta_s = -0.308(30)$ from different sets of BCs, clearly
smaller in modulus than previous estimates
\cite{morris:86a,kawamura:91,maucourt:98a,kosterlitz:99a}, and rather close to
$\theta_s = -0.28$ found for the {\em Gaussian\/} 2D Ising case. The chiral exponent
$\theta_c=-0.114(16)$, on the other hand, is found to be clearly different from
$\theta_s$ and closer to $\theta_s = 0$ found for the {\em bimodal\/} 2D Ising spin
glass.  Note also, that our results are rather far from $\theta_s = -1/\nu_s = -1.0$
resp.\ $\theta_c = -1/\nu_c = -0.5$ estimated by finite-temperature Monte Carlo
simulations \cite{ray:91a}, which probably is due to equilibration problems at low
temperatures.

\ack

The research at the University of Waterloo was undertaken, in part, thanks to funding
from the Canada Research Chairs Program (M.G.). M.W.\ acknowledges support by the EC
under contract No.\ MEIF-CT-2004-501422.

\end{document}